\documentclass[aps,prc,twocolumn,groupedaddress,showpacs]{revtex4-1}
\usepackage{graphicx}
\usepackage{comment}

\begin{document}


\title{$\mathbf{\emph{K}^{*}(892)^{+}}$ production in proton-proton collisions at $\mathbf{E_{beam} = 3.5}$~GeV}



\author{G.~Agakishiev$^{7}$, O.~Arnold$^{10,9}$, D.~Belver$^{18}$, A.~Belyaev$^{7}$, 
J.C.~Berger-Chen$^{10,9}$, A.~Blanco$^{2}$, M.~B\"{o}hmer$^{10}$, J.~L.~Boyard$^{16}$, P.~Cabanelas$^{18}$, 
S.~Chernenko$^{7}$, A.~Dybczak$^{3}$, E.~Epple$^{10,9}$, L.~Fabbietti$^{10,9}$, O.~Fateev$^{7}$, 
P.~Finocchiaro$^{1}$, P.~Fonte$^{2,b}$, J.~Friese$^{10}$, I.~Fr\"{o}hlich$^{8}$, T.~Galatyuk$^{5,c}$, 
J.~A.~Garz\'{o}n$^{18}$, R.~Gernh\"{a}user$^{10}$, K.~G\"{o}bel$^{8}$, M.~Golubeva$^{13}$, D.~Gonz\'{a}lez-D\'{\i}az$^{5}$, 
F.~Guber$^{13}$, M.~Gumberidze$^{5,c}$, T.~Heinz$^{4}$, T.~Hennino$^{16}$, R.~Holzmann$^{4}$, 
A.~Ierusalimov$^{7}$, I.~Iori$^{12,e}$, A.~Ivashkin$^{13}$, M.~Jurkovic$^{10}$, B.~K\"{a}mpfer$^{6,d}$, 
T.~Karavicheva$^{13}$, I.~Koenig$^{4}$, W.~Koenig$^{4}$, B.~W.~Kolb$^{4}$, G.~Korcyl$^{3}$, 
G.~Kornakov$^{5}$, R.~Kotte$^{6}$, A.~Kr\'{a}sa$^{17}$, F.~Krizek$^{17}$, R.~Kr\"{u}cken$^{10}$, 
H.~Kuc$^{3,16}$, W.~K\"{u}hn$^{11}$, A.~Kugler$^{17}$, T.~Kunz$^{10}$, A.~Kurepin$^{13}$, 
V.~Ladygin$^{7}$, R.~Lalik$^{10,9}$, K.~Lapidus$^{10,9,\ast}$, A.~Lebedev$^{14}$, L.~Lopes$^{2}$, 
M.~Lorenz$^{8,h}$, L.~Maier$^{10}$, A.~Mangiarotti$^{2}$, J.~Markert$^{8}$, V.~Metag$^{11}$, 
J.~Michel$^{8}$, D.~Mihaylov$^{10,9,\ast}$, C.~M\"{u}ntz$^{8}$, R.~M\"{u}nzer$^{10,9}$, L.~Naumann$^{6}$, Y.~C.~Pachmayer$^{8}$, 
M.~Palka$^{3}$, Y.~Parpottas$^{15,f}$, V.~Pechenov$^{4}$, O.~Pechenova$^{8}$, J.~Pietraszko$^{4}$, 
W.~Przygoda$^{3}$, B.~Ramstein$^{16}$, A.~Reshetin$^{13}$, A.~Rustamov$^{8}$, A.~Sadovsky$^{13}$, 
P.~Salabura$^{3}$, A.~Schmah$^{a}$, E.~Schwab$^{4}$, J.~Siebenson$^{10,9}$, Yu.G.~Sobolev$^{17}$, S.~Spataro$^{g}$, B.~Spruck$^{11}$, H.~Str\"{o}bele$^{8}$, J.~Stroth$^{8,4}$, C.~Sturm$^{4}$, O.~Svoboda$^{17}$, A.~Tarantola$^{8}$, 
K.~Teilab$^{8}$, P.~Tlusty$^{17}$, M.~Traxler$^{4}$, H.~Tsertos$^{15}$, T.~~Vasiliev$^{7}$, 
V.~Wagner$^{17}$, M.~Weber$^{10}$, C.~Wendisch$^{4}$, J.~W\"{u}stenfeld$^{6}$, S.~Yurevich$^{4}$, 
Y.~Zanevsky$^{7}$}

\affiliation{
(HADES collaboration) \\\mbox{$^{1}$Istituto Nazionale di Fisica Nucleare - Laboratori Nazionali del Sud, 95125~Catania, Italy}\\
\mbox{$^{2}$LIP-Laborat\'{o}rio de Instrumenta\c{c}\~{a}o e F\'{\i}sica Experimental de Part\'{\i}culas, 3004-516~Coimbra, Portugal}\\
\mbox{$^{3}$Smoluchowski Institute of Physics, Jagiellonian University of Cracow, 30-059~Krak\'{o}w, Poland}\\
\mbox{$^{4}$GSI Helmholtzzentrum f\"{u}r Schwerionenforschung GmbH, 64291~Darmstadt, Germany}\\
\mbox{$^{5}$Technische Universit\"{a}t Darmstadt, 64289~Darmstadt, Germany}\\
\mbox{$^{6}$Institut f\"{u}r Strahlenphysik, Helmholtz-Zentrum Dresden-Rossendorf, 01314~Dresden, Germany}\\
\mbox{$^{7}$Joint Institute of Nuclear Research, 141980~Dubna, Russia}\\
\mbox{$^{8}$Institut f\"{u}r Kernphysik, Goethe-Universit\"{a}t, 60438 ~Frankfurt, Germany}\\
\mbox{$^{9}$Excellence Cluster 'Origin and Structure of the Universe', 85748~Garching, Germany}\\
\mbox{$^{10}$Physik Department E12, Technische Universit\"{a}t M\"{u}nchen, 85748~Garching, Germany}\\
\mbox{$^{11}$II.Physikalisches Institut, Justus Liebig Universit\"{a}t Giessen, 35392~Giessen, Germany}\\
\mbox{$^{12}$Istituto Nazionale di Fisica Nucleare, Sezione di Milano, 20133~Milano, Italy}\\
\mbox{$^{13}$Institute for Nuclear Research, Russian Academy of Science, 117312~Moscow, Russia}\\
\mbox{$^{14}$Institute of Theoretical and Experimental Physics, 117218~Moscow, Russia}\\
\mbox{$^{15}$Department of Physics, University of Cyprus, 1678~Nicosia, Cyprus}\\
\mbox{$^{16}$Institut de Physique Nucl\'{e}aire (UMR 8608), CNRS/IN2P3 - Universit\'{e} Paris Sud, F-91406~Orsay Cedex, France}\\
\mbox{$^{17}$Nuclear Physics Institute, Academy of Sciences of Czech Republic, 25068~Rez, Czech Republic}\\
\mbox{$^{18}$LabCAF. F. F\'{\i}sica, Univ. de Santiago de Compostela, 15706~Santiago de Compostela, Spain}\\ 
\\
\mbox{$^{a}$ also at Lawrence Berkeley National Laboratory, ~Berkeley, USA}\\
\mbox{$^{b}$ also at ISEC Coimbra, ~Coimbra, Portugal}\\
\mbox{$^{c}$ also at ExtreMe Matter Institute EMMI, 64291~Darmstadt, Germany}\\
\mbox{$^{d}$ also at Technische Universit\"{a}t Dresden, 01062~Dresden, Germany}\\
\mbox{$^{e}$ also at Dipartimento di Fisica, Universit\`{a} di Milano, 20133~Milano, Italy}\\
\mbox{$^{f}$ also at Frederick University, 1036~Nicosia, Cyprus}\\
\mbox{$^{g}$ also at Dipartimento di Fisica and INFN, Universit\`{a} di Torino, 10125~Torino, Italy}\\
\mbox{$^{h}$ also at Utrecht University, 3584 CC~Utrecht, The Netherlands}\\
\\
\mbox{$^{\ast}$ corresponding authors: kirill.lapidus@ph.tum.de, dimitar.mihaylov@mytum.de}
}


\date{\today}

\begin{abstract}
We present results on the $K^{*}(892)^{+}$ production in proton-proton collisions at a beam energy of $E = 3.5$~GeV,
which is hitherto the lowest energy at which this mesonic resonance has been observed in nucleon-
nucleon reactions. The data are interpreted within a two-channel model
that includes the 3-body production of $K^{*}(892)^{+}$ associated with the $\Lambda$- or $\Sigma$-hyperon. The relative contributions of both channels are estimated. Besides the total
cross section $\sigma(p+p \to K^{*}(892)^{+} + X) = 9.5 \pm 0.9 ^{+1.1}_{-0.9} \pm 0.7$~$\mu$b, that adds a new data point to the excitation function of the $K^{*}(892)^{+}$ production in the region of low excess energy, transverse momenta and angular spectra are extracted and compared with the predictions of the two-channel model. The spin characteristics of $K^{*}(892)^{+}$ are discussed as well in terms of the spin-alignment.
\end{abstract}

\pacs{25.75.Dw, 14.40.Df}

\maketitle


\section{Introduction}

The production of the strange vector meson $K^{*}(892)$ in proton-proton collisions was studied rather extensively
at high collision energies \cite{Ammosov:1975cn, Kichimi:1979cs, Drijard:1981ab, Brick:1982ke, Aziz:1985gf, Bogolyubsky:1988ei, AguilarBenitez:1991yy}, with the lowest-energy data point at $\sqrt{s} = 4.93$~GeV \cite{Bockmann:1979fr}. No data, however, exist in the region close to the production threshold $\sqrt{s_{thr}} = 2.95$~GeV. This is in contrast to the situation with the kaon ground state, the production of which has been measured both
inclusively and exclusively in a broad range of energies,
including the very vicinity of its production threshold.
The production of kaons and their excitations in proton-proton collisions is
governed by the conservation of strangeness, so the simplest
possible reaction reads
\begin{equation}
p + p \to N + Y + K/K^{*}(892),
\end{equation}
where $N$ stands for the nucleon and $Y$ for the ground-state
hyperons $\Lambda(1116)$ or $\Sigma(1189)$. The 3-body kaon production has been studied in detail at
low excess energies \cite{AbdelBary:2010pc, AbdelBary:2012zz}, and it was established that, for
energies close to the production threshold, the kaon production
is mainly accompanied by a $\Lambda(1116)$-hyperon
rather than a $\Sigma(1189)$. It is of interest, therefore, to identify
the preferable formation mechanism of the $K^{*}$-meson
as well.

In this paper we discuss the study of the $K^{*}(892)^{+}$
production in proton-proton collisions performed by the
HADES collaboration. The deep sub-threshold $K^{*}$ production was analyzed in \cite{Agakishiev:2013nta} for
Ar+KCl reactions at a beam energy of 1.756 GeV. In view of future experiments at the Facility
for Antiproton and Heavy-Ion Research (FAIR) exploring
heavy ion collisions at energies of 2-8 GeV/nucleon, new
data from proton-proton reactions are essential as reference
measurements and input for transport models.
This work complements our previous studies of inclusive and
exclusive strangeness production in proton-proton reactions
at 3.5 GeV, namely $K^{0}$ \cite{Agakishiev:2013yyy}, $\Sigma(1385)^{+}$ \cite{Agakishiev:2011qw},
$\Lambda(1405)$ \cite{Agakishiev:2012qx, Agakishiev:2012xk}, and $pK^{+}\Lambda$ \cite{Agakishiev:2014dha}.
The paper is organized as follows. Section II gives
a brief information about the experimental setup. The
particle selection and $K^{*}$ reconstruction procedure is described
in Section III. Section IV contains the obtained results,
their interpretation within the two-channel model,
and a discussion of the spin alignment measurement. The summary can be found in Section V.

\section{The experiment}

The experimental data stem from the High-Acceptance
Di-Electron Spectrometer (HADES), installed at the
SIS18 synchrotron (GSI Helmholtzzentrum, Darmstadt).
The detector tracking system consists of a superconducting
magnet and four planes of Multiwire Drift Chambers
(MDC). The particle identification capabilities are extended
by the Time-of-Flight wall and a Ring Imaging
Cherenkov (RICH) detector. The detector, as implied
by its name, is characterized by a large acceptance both
in the polar (from $18^{\circ}$ to $85^{\circ}$) and azimuthal angles.
The detector sub-systems are described in detail in \cite{Agakishiev:2009am}.
In 2007 a measurement of proton-proton collisions at a
kinetic beam energy of 3.5 GeV was performed: the beam
with an average intensity of about $1 \times 10^{7}$ particles/s was
incident on a liquid hydrogen target with an area density of 0.35~g$/$cm$^{2}$ corresponding to total interaction probability of $\sim 0.7 \%$.
In total, $1.2 \times 10^{9}$ events were collected. The first-level trigger (LVL1)
required at least three hits in the Time-of-Flight
wall in order to suppress elastic scattering events.

\section{Data analysis}
\subsection{$K^{*}$ reconstruction}
\begin{figure}[h]
\includegraphics[width=0.39\textwidth, angle=90]{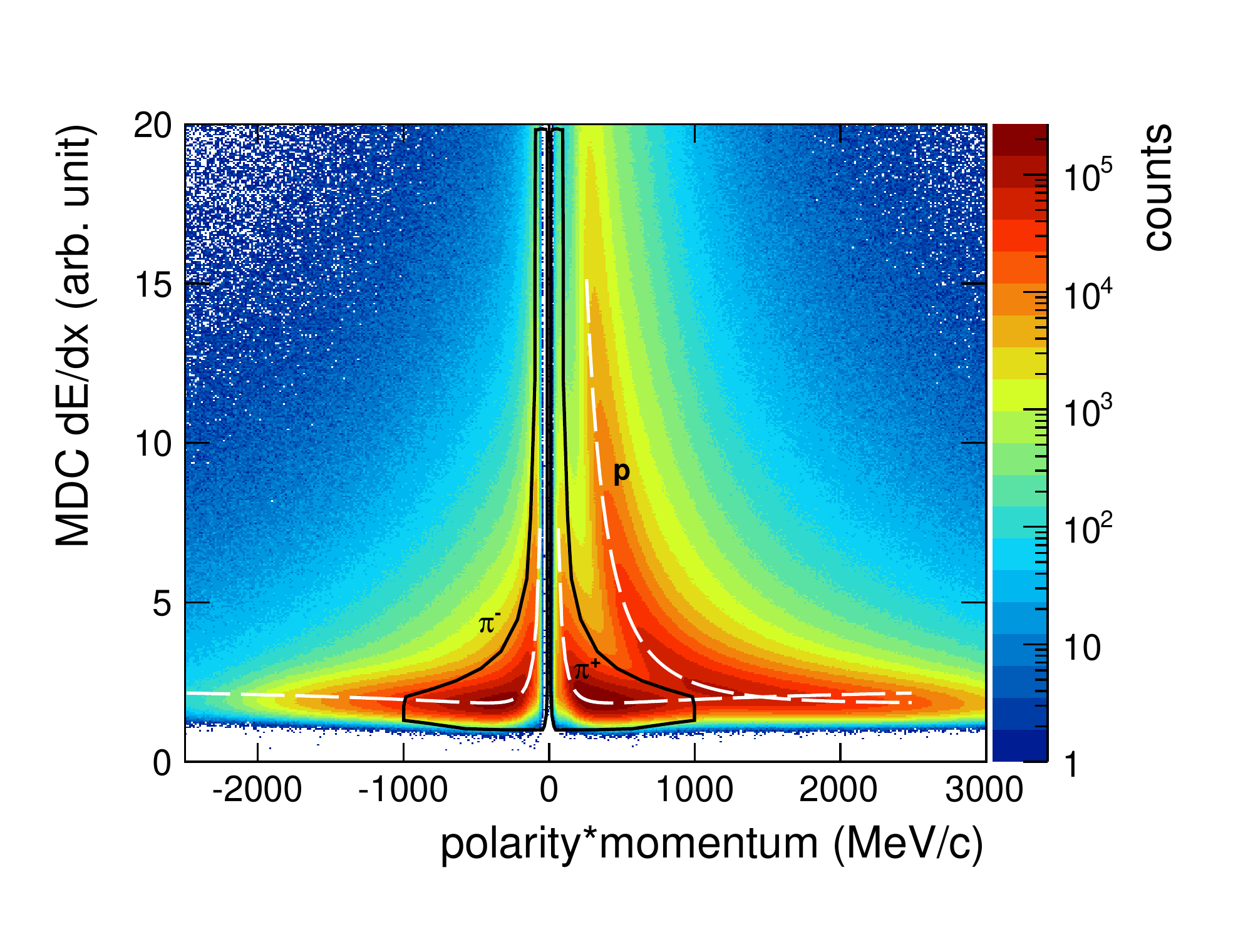}
\caption{\label{fig:dEdx} (Color online) Specific energy loss of charged particles in MDC 
chambers as a function of the momentum times polarity. Solid curves show graphical cuts for the $\pi^{\pm}$ selection. The Bethe-Bloch curves are shown by dashed curves.}
\end{figure}

The $K^{*}(892)^{+}$-meson decays strongly, at the primary
pp-reaction vertex, into a kaon-pion pair. The decay
mode $K^{*}(892)^{+} \to K^{0} \pi^{+}$ with a branching ratio of 2/3 is particularly suited for the analysis, since the short-lived
component of the $K^{0}$, the $K^{0}_{S}$, decays weakly into
a $\pi^{+}\pi^{-}$ pair ($c\tau = 2.68$~cm, branching ratio 69.2\%). The
considered final state of the $K^{*}(892)^{+}$ decay is, therefore,
composed out of three charged pions, two of which are emitted
from a secondary vertex. Therefore, as a first step of
the analysis, we select events with (at least) three pions,
identified by a selection on the $(dE/dx)_{MDC}$-momentum
plane (Fig.~\ref{fig:dEdx}). The variable $(dE/dx)_{MDC}$ is the cumulative
specific energy loss of a charged particle in the four
MDC planes.

\begin{figure}[h]
\includegraphics[width=0.48\textwidth]{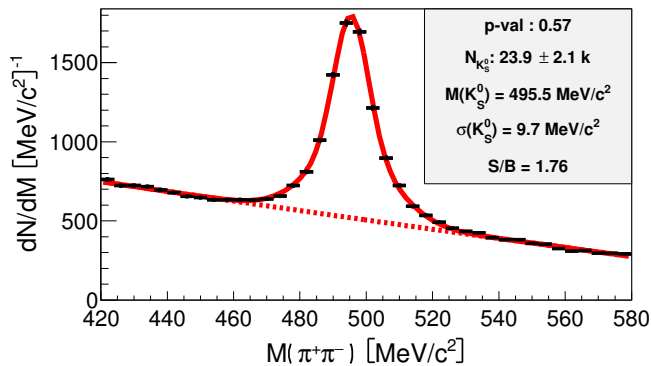}
\caption{\label{fig:K0S} (Color online) $\pi^{+}\pi^{-}$ invariant mass distribution from events with at least three pions used for the $K^{0}_S$ reconstruction. The legend delivers information on the fit p-value, the number of $K^{0}_{S}$ over background, extracted mass, width, and signal to background ratio.}
\end{figure}

In the next step we consider all triplets of charged pions ($\pi^{+}\pi^{+}\pi^{-}$) 
that were found in one event. Since two positively
charged pions are available, two $K^{0}_S$-candidates are
constructed for each triplet. Afterwards, to ensure that
the $K^{0}_S$ decayed away from the primary vertex, and, thus,
reduce the combinatorial background, a set of topological
cuts was applied to the $\pi^{+}\pi^{-}$ pair that forms the $K^{0}_S$ candidate.
These were: i) a cut on the distance between the primary and the secondary vertex $d(K^{0}_S - V)>28$~mm, ii) a
cut on the distance of closest approach between two pion
tracks, $d_{\pi^{+}-\pi^{-}}<13$~mm, and iii) a cut on the distance of
closest approach between either of the extrapolated pion
tracks and the primary vertex $DCA^{K^{0}}_{\pi} > 8$~mm. Besides,
a cut on the distance of closest approach between
the pion track \emph{not associated} to the $K^{0}_S$-candidate (i.e.
stemming from the $K^{*}$-decay) and the primary vertex $DCA^{K^{*}}_{\pi} < 6$~mm, has been introduced. The application
of the topological cuts reduces the double counting probability
(i.e. identifying more than one $K^{0}_S$-candidate in
one event, which is kinematically forbidden at this collision
energy) to 6\%.
The invariant mass spectrum of the  $\pi^{+}\pi^{-}$ pairs that
passed the topological cuts is shown in Fig.~\ref{fig:K0S}. A prominent $K^{0}_S$ signal is visible. In total, about $24 \times 10^{3}$ $K^{0}_S$-candidates were reconstructed in events with at least three charged pions.

\begin{figure*}[t]
\includegraphics[width=0.88\textwidth]{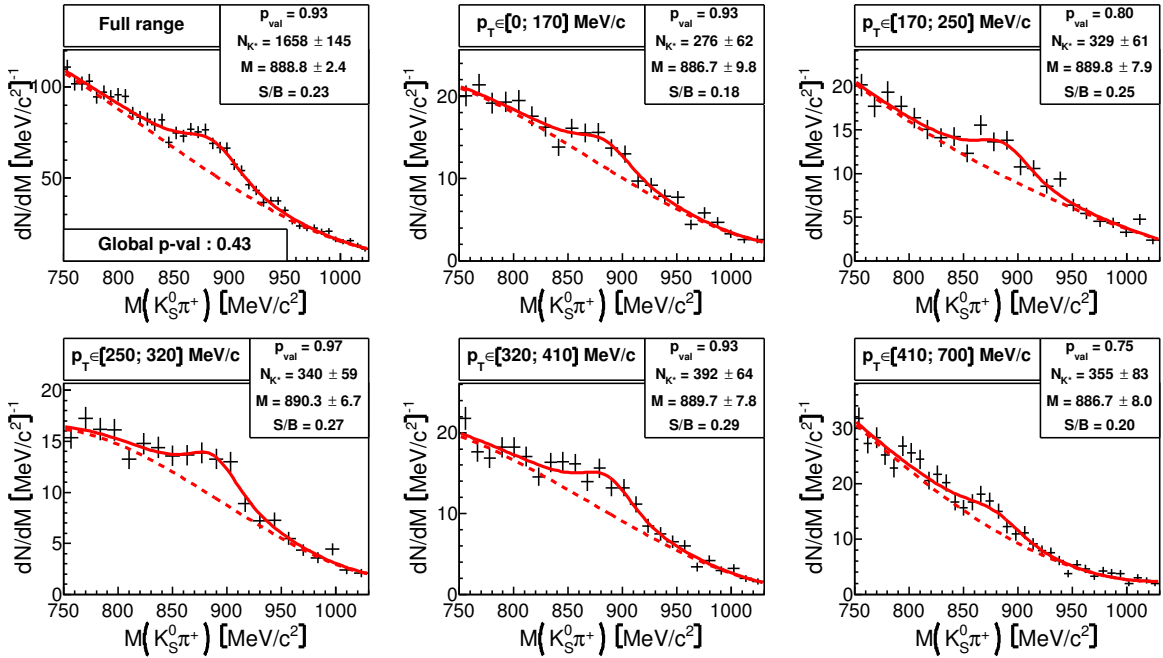}
\caption{\label{fig:KstarIM} (Color online) Invariant mass spectra of $K^{0}_S-\pi^{+}$ pairs (symbols) for the total sample (upper left) and in five transverse momentum bins. Solid curves are for the fits with \ref{eq:fit_fun}, dashed curves depict the background, separately. The boxes deliver information on the fit p-value, the number of $K^{*}$ over background, mean mass and signal to background ratio.}
\end{figure*}

In the next step we applied a cut on the invariant mass
of the $\pi^{+}\pi^{-}$ pairs, i.e. a 19.4~MeV$/c^{2}$ interval ($\pm 1\sigma$) centered
at the $K^{0}_S$ mass peak at 495.5~MeV$/c^{2}$ and combined the pairs passing the cut with the remaining positively
charged pion. The resulting invariant mass distributions of the $K^{0}_{S}-\pi^{+}$ system are shown in Fig.~\ref{fig:KstarIM} for the total sample (upper left panel)
and separately in five transverse momentum bins. A clear signal of the sought-after $K^{*}$ is visible on top of the combinatorial background
that can be divided into two classes: i)  $\pi^{+}\pi^{+}\pi^{-}$ triplets
produced in reactions without strangeness involvement, and ii) non-resonant production of $K^{0}_{S}\pi^{+}$ pairs.
The total statistics of about 1700 $K^{*}$ allows for a one-dimensional differential analysis. Below we discuss the
extraction of the transverse momentum spectra in detail;
the procedure can be applied to any kinematical variable.
The extraction of the raw (i.e. neither corrected for the limited geometrical acceptance of the
HADES detector nor for the efficiency of the analysis procedure) $K^{*}$ yields is carried out with
fits of the experimental invariant mass distributions. After a careful study of the best way to
approximate the experimental distributions, we used the following function
\begin{equation}
\label{eq:fit_fun}
f \left( M \right) = F_{PS}\left(M\right) \times V(M;\Gamma,\sigma) + P_{3}\left(M\right),
\end{equation}
where $M$ is the invariant mass of $K^{0}_{S}\pi^{+}$ pairs, $F_{PS}\left(M\right)$ --- the factor that takes into account phase-space limitations, $V\left(M; \Gamma,\sigma \right)$ --- the Voigt function, and $P_{3}\left(M\right)$ --- a third degree polynomial that models the non-resonant background. The parameters of the Voigt function are: i) $\Gamma$ --- internal width of $K^{*}$ that was fixed to the PDG value
of 50.8~MeV \cite{Agashe:2014kda} and ii) $\sigma$ --- the detector resolution (instrumental
width) that was determined (and fixed as well) to
be about $11$~MeV by simulating a sample of zero-width $K^{*}$'s.
The phase space distortion factor $F_{PS}\left(M\right)$ depends on
the production channel ($\Lambda$- or $\Sigma$-associated), so first we
shall introduce the two-channel simulation model.

\subsection{Two channel model}

Due to the quite low beam energy, the three following 3-body channels are expected to dominate $K^{*}$ production:
\begin{equation}
\label{eq:lch}
p + p \to p + \Lambda + K^{*}(892)^{+},
\end{equation}
\begin{equation}
\label{eq:sch}
p + p \to p + \Sigma^{0} + K^{*}(892)^{+},
\end{equation}
\begin{equation}
p + p \to n + \Sigma^{+} + K^{*}(892)^{+}.
\end{equation}
Here we neglect the contribution of the 4-body channels
with an additional pion ($p+p \to N + \pi + Y + K^{*}$). They
are energetically possible, but are expected to be suppressed in
comparison with the 3-body channels. As it will be shown below, this assumption is confirmed by the experimental
data.
We make another simplification, namely we consider only
one $\Sigma$-associated channel ($p\Sigma^{0}K^{*}$) out of two that are
allowed. Up to the small differences in the masses of
the reaction products, both channels have exactly the
same kinematics and are indistinguishable in the inclusive
analysis of $K^{*}$. Hence, we employ hereafter a two-channel
model that includes the $\Lambda$- and the $\Sigma$-channels,
where the latter has to be understood as the sum of
the two isospin-splitted sub-channels.
The invariant mass distributions of the simulated $K^{*}$'s
were reconstructed for both channels of the model. Then, the phase-space factor $F_{PS}\left(M\right)$ was determined,
which takes into account the deviation from an ideal Breit-Wigner shape. As the contributions of both channels are
not known a priori, the phase-space factor is constructed
as the sum of the two individual contributions from the $\Lambda$- and $\Sigma$-channel:
\begin{equation}
F_{PS} \left(M \right) = A_{\Lambda} \times F^{\Lambda}_{PS} \left( M \right) +  A_{\Sigma} \times F^{\Sigma}_{PS} \left( M \right).
\end{equation}
It was found that the fitting procedure is not sensitive to
the exact contribution of each channel (i.e to the weights $A_{\Lambda}$ and $A_{\Sigma}$).

\subsection{Raw spectra and corrections}

\begin{figure}[h]
\includegraphics[width=0.49\textwidth]{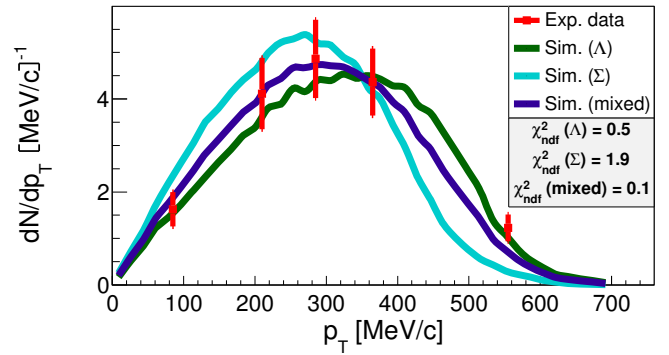}
\caption{\label{fig:pt_raw} (Color online) Raw $p_{t}$-spectrum of $K^{*}$'s produced in proton-proton
reactions. Markers --- experimental data with statistical uncertainties, curves --- expectations
from the channels (3), (4), and their mixture ``$0.6 \times \Lambda + 0.4 \times \Sigma$''.}
\end{figure}

Fits of the experimental data, performed with the Eq.~(\ref{eq:fit_fun}) allow to extract the raw yields (affected by the finite acceptance of the detector and efficiency of the analysis
procedure). As an example, the raw $p_{t}$-spectrum of
the $K^{*}$ mesons is shown in Fig.~\ref{fig:pt_raw}. Also shown are $K^{*}$ distributions corresponding to the $\Lambda$- and $\Sigma$-channels of the two-channel model. They were obtained in the following
way. A set of events (four-vectors of the reactions products) corresponding to both channels (\ref{eq:lch}) and (\ref{eq:sch})
were simulated with the {\sc pluto} Monte Carlo generator \cite{Frohlich:2007bi}. A uniform population of the 3-body phase-space
has been assumed: neither any kind of angular anisotropy has been implemented, nor any contribution from
an $N^{*}$- or $\Delta$-resonance coupling to the $K^{*}-Y$ pair considered.
Afterwards, these events served as input for the
full-scale simulation procedure that includes the propagation
of particles in the detector (the {\sc geant3} code
was employed), track reconstruction, etc. Finally, the
simulated data sample was analyzed in the same way as
the experimental one. All simulated curves shown in
Fig.~\ref{fig:pt_raw} are normalized to the integral of the experimental
distribution. Furthermore, an optimal mixture of the two channels has been determined by means of a $\chi^{2}$-analysis, delivering the value for the relative $\Sigma$-channel contribution of $0.4\pm0.2$, where $0.2$ is the dominating systematic uncertainty determined by the variation of the experimental cuts, as will be explained below. The resulting mixed spectrum is shown in Fig.~\ref{fig:pt_raw} as well.
A contribution of the 4-body channels ($N\pi Y K^{*}$ final state) would produce an even
softer $p_{t}$-spectrum as compared to the one generated by the $\Sigma$-channel and, therefore, is completely negligible in this analysis. Finally, we note that an analysis of the missing mass to the $K^{*}p$ final state, potentially more selective with respect to the $\Lambda$- or $\Sigma$-contributions, is not feasible due to the limited statistics.

As the constructed model (``$0.6 \times \Lambda + 0.4 \times \Sigma$'') describes the data very well, we
can use it to correct the raw experimental yields. Due to the limited statistics of the experimental
data we perform a one-dimensional correction. The efficiency and acceptance depend, thus, on one kinematical
variable. We exemplify the procedure for the $p_{t}$-variable; all other variables are treated analogously.

For the purpose of the efficiency correction we prepare
a histogram $I$, which corresponds to the $p_{t}$-spectrum of
the simulated $K^{*}$'s in the full solid angle, not affected by the limited
geometrical acceptance of the detector and efficiency
of the analysis procedure. A histogram $O$ corresponds to
the $p_{t}$-spectrum of $K^{*}$'s that went through full simulation
and analysis chain. Finally, the ratio $\epsilon(p_{t}) = O/I$ is
the efficiency histogram. Dividing bin-wise the raw experimental $p_{t}$-spectrum by the $\epsilon(p_{t})$ histogram we obtain the acceptance- and efficiency-corrected spectrum.

\section{Results and discussion}

\begin{figure}[h]
\includegraphics[width=0.5\textwidth]{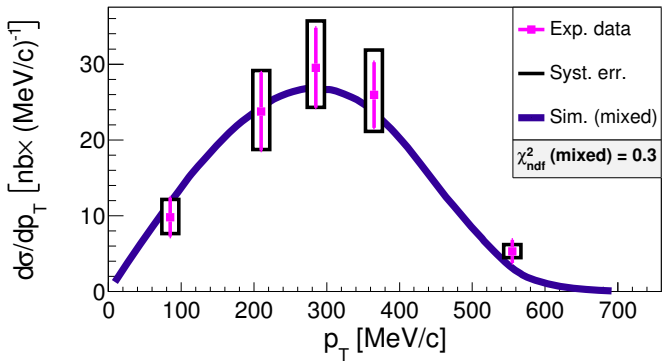}
\caption{\label{fig:pt_corr} (Color online) $K^{*}$ transverse momentum spectrum. Markers --- experimental data with statistical uncertainties, empty boxes --- systematical uncertainties. The two-channel model (``$0.6 \times \Lambda + 0.4 \times \Sigma$'') final state phase space distribution is shown by the solid curve.}
\end{figure}

Figure~\ref{fig:pt_corr} shows the acceptance and efficiency corrected $p_{t}$-spectrum of $K^{*}$. The
experimental data are normalized absolutely based on
the analysis of the proton-proton elastic scattering channel in the
HADES acceptance, as has been done for the inclusive
di-electron analysis \cite{HADES:2011ab}. To no surprise (cf. Fig.~\ref{fig:pt_raw}),
the corrected spectrum is very well described by the two-channel
model that assumes a 40\% contribution of the $\Sigma$-channel.

\begin{figure}[h]
\includegraphics[width=0.5\textwidth]{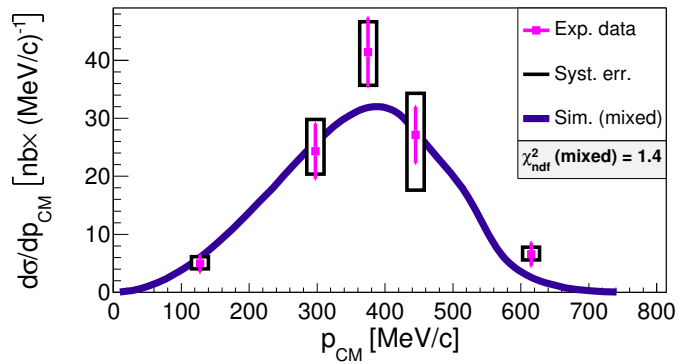}
\caption{\label{fig:ptot_corr} (Color online) Same as Fig.~\ref{fig:pt_corr} for the momentum spectrum in the
pp centre-of-mass reference frame.}
\end{figure}

\begin{figure}[h]
\includegraphics[width=0.5\textwidth]{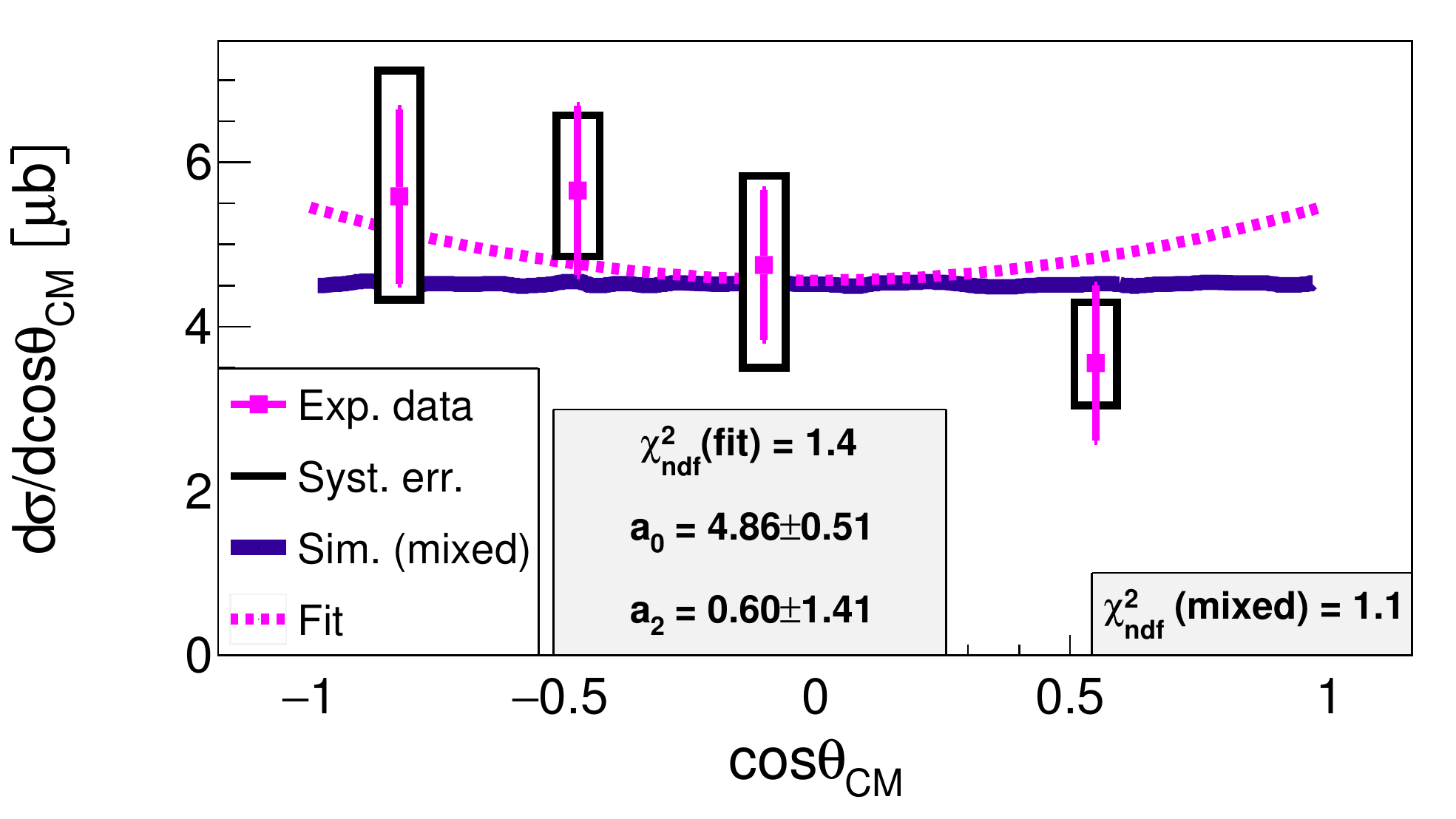}
\caption{\label{fig:cost_corr} (Color online) Same as Fig.~\ref{fig:pt_corr} for the angular spectrum in the
pp-centre-of-mass reference frame. The dashed curve correspond to the fit with Legendre polynomials (only 0-th and 2-nd order are used due to the symmetry arguments, resulting coefficients are shown in the inset).}
\end{figure}

As mentioned already, a one-dimensional
analysis, exemplified above with the transverse momentum
variable, can be performed for any chosen kinematical
variable. For completeness, Figs.~\ref{fig:ptot_corr} and \ref{fig:cost_corr} show corrected
spectra for the total momentum and angular distribution in the
pp centre-of-mass reference frame, respectively.

Remarkably, all three selected ($p_{t}$, $p_{c.m.}$, and $\cos{\theta_{c.m.}}$) projections of a three-dimensional single-particle phase
space are well described by the two-channel model. No significant angular anisotropy is observed for the $K^{*}$ emission in proton-proton
collisions: a Legendre-polynomial fit (dotted curve in Fig.~\ref{fig:cost_corr}) does not deliver a better $\chi^{2}$ as compared to the isotropic (by construction) distribution of the two-channel model. 
The integration of the experimental $p_{t}$-spectrum
(Fig.~\ref{fig:pt_corr}) allows to extract the total cross section of the
inclusive $K^{*+}$ production:
\begin{equation}
\label{eq:sigma_tot}
\sigma(p+p \to K^{*}(892)^{+} + X) = 9.5 \pm 0.9 ^{+1.1}_{-0.9} \pm 0.7~\mu\text{b},
\end{equation}
where the statistical (first), systematic (second) and normalization
(third) uncertainties are given.
The systematic uncertainty was estimated by a variation
of the experimental cuts (topological cuts plus $K^{0}_S$ selection via an invariant mass constraint). The values of
the cuts used for this variations are listed in Table~\ref{table:table1}. In
total, 1200 cut combinations have been tested. For each
cut combination new invariant mass fits were performed
along with new efficiency corrections. Afterwards, from
the distribution of the total cross section a central interval covering
68\% of the outcomes has been identified. The borders
of this interval---asymmetric with respect to the median
value---define the systematic uncertainty quoted in
Eq.~\ref{eq:sigma_tot}.

\begin{table}[h]
\caption{\label{table:table1}%
Topological and the $K^{0}_S$ invariant mass cut variations
used to estimate the systematic uncertainty. For all
cut combinations the condition $DCA^{K^{*}}_{\pi} \leq DCA^{K^{0}}_{\pi}$ has been
demanded, reducing 1800 cut combinations to 1200.}
\begin{ruledtabular}
\begin{tabular}{llll}
Observable & Lower value & Upper value & Steps \\
\colrule
$d(K^{0}_S - V)$~[mm] & $> 24$ & $> 40$ & 5\\
$d_{\pi^{+}-\pi^{-}}$~[mm] & $<7$ & $<13$ & 4\\
$DCA^{K^{0}}_{\pi}$~[mm] & $>5.6$ & $>16$ & 5\\
$DCA^{K^{*}}_{\pi}$~[mm] & $<3$ & $<16$ & 6\\
\colrule
$\Delta M_{K^{0}_S}$~[MeV/$c^{2}$]& 9.7(1$\sigma$) & 19.4(2$\sigma$) & 3\\
\end{tabular}
\end{ruledtabular}
\end{table}

\begin{figure}[h]
\includegraphics[width=0.52\textwidth]{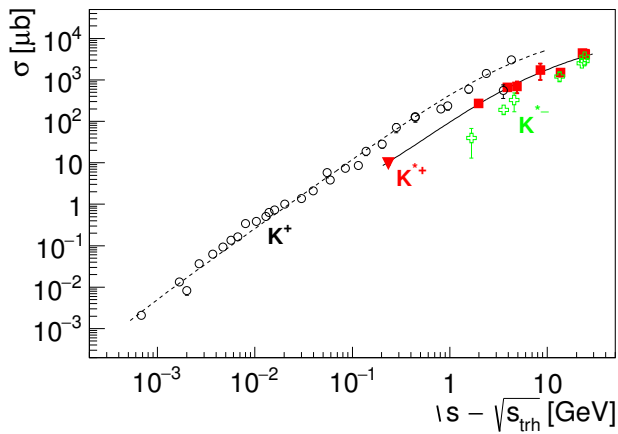}
\caption{\label{fig:cs} (Color online) Energy ($\sqrt{s} - \sqrt{s_{thr}}$) dependence of the total cross section for the
processes: i) $pp \to K^{*}(892)^{+}X$ (red squares --- world data \cite{Ammosov:1975cn, Bockmann:1979fr, Bogolyubsky:1988ei, Brick:1982ke, Aziz:1985gf, AguilarBenitez:1991yy, Kichimi:1979cs}, red triangle --- present work), ii) $pp \to K^{*}(892)^{-}X$ (empty green crosses) \cite{Bockmann:1979fr, Bogolyubsky:1988ei, Brick:1982ke, Aziz:1985gf, AguilarBenitez:1991yy, Kichimi:1979cs}, and iii) $pp \to K^{+}X$ (empty circles) (\cite{AbdelBary:2010pc, AbdelBary:2012zz, Reed:1968zza} and references therein). The solid (dashed) line is a fit to the $K^{*}(892)^{+}$ ($K^{+}$) data with $f\left( x \right) = C \left( 1 - \left(D/x \right)^{\mu} \right)^{\nu}$, where $x = \sqrt{s}$. The numerical values are $C = 3.22 \times 10^{6} (1.04 \times 10^{5})$, $D = 2.89 (2.55)$~GeV, $\mu = 1.19 \times 10^{-2} (1.16 \times 10^{-1})$, $\nu = 1.86 (1.67)$. }
\end{figure}

The extracted total cross section value complements
the $K^{*}$ excitation function shown in Fig.~\ref{fig:cs} in the low-energy
region, where measurements were not available until now. The HADES data point is consistent with the trend set by the measurements at higher energies.

\begin{figure}[h]
\includegraphics[width=0.485\textwidth]{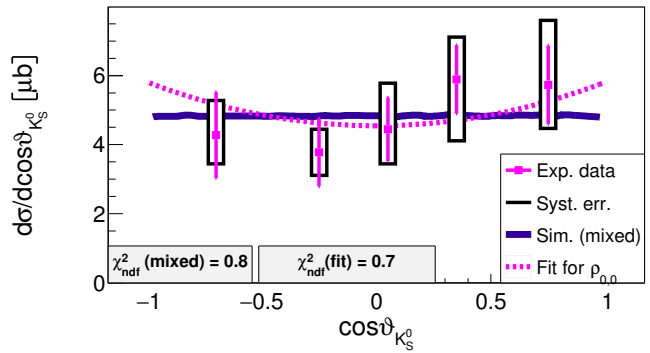}
\caption{\label{fig:spinal} (Color online) Angular distribution of $K^{0}_S$ in the $K^{*}$ rest frame.
Markers --- experimental data with statistical uncertainties,
empty boxes --- systematical uncertainties. The two-channel-model (``$0.6 \times \Lambda + 0.4 \times \Sigma$'') final state phase space distribution as simulated with {\sc pluto} without any spin alignment is shown by the solid curve. 
The dashed curve --- fit of Eq. (\ref{eq:spinalfit}) to the data with $\rho_{00} = 0.39$.}
\end{figure}

In comparison to the ground state, the $K^{*}$ carries spin
one, so we proceed with the discussion of its polarisation
properties as probed in proton-proton collisions. The
spin configuration of the $K^{*}$ in the final state is described
by the spin density matrix $\rho_{mm'}$. The diagonal elements
$\rho_{11}$, $\rho_{00}$, and $\rho_{1-1}$ define, respectively, the probabilities
of the $+1$, 0 and $-1$ spin projections on the quantisation
axis. 
The $\rho_{00}$ can be 
extracted from the
angular distribution of the decay products ($K^{0}$ or $\pi^{+}$)
in the rest frame of $K^{*}$ \cite{Donoghue:1978bx} via
\begin{equation}
\label{eq:spinalfit}
W \left(\vartheta \right) = \frac{3}{4} \left[ \left(1 - \rho_{00} \right) + \left(3\rho_{00} - 1\right)\cos^{2}(\vartheta)\right].
\end{equation}
The situation with $\rho_{00} \neq 1/3$ is referred to as the \emph{spin-
alignment} case, i.e. not equally probable populations of
the $\pm 1$ and 0 spin projections.
The angular distribution of interest is shown in Fig.~\ref{fig:spinal}.
Our fit with Eq.~(\ref{eq:spinalfit}) gives the following result:
\begin{equation}
\rho_{00} = 0.39 \pm 0.09(\text{stat.})^{+0.10}_{-0.09}(\text{syst.}).
\end{equation}
Within uncertainties our measurement is fully consistent
with $\rho_{00} =1/3$, i.e. no spin-alignment of $K^{*}$ is
observed.

\section{Summary and conclusions}
To summarize, we presented here the hitherto lowest-energy
measurement of the $K^{*}(892)^{+}$ production in proton-proton collisions. The relative contribution of the channel $p + p \to p + \Sigma + K^{*}(892)^{+}$ has been estimated as $0.4\pm0.2$. 
Within uncertainties of the experimental data, no deviations from a 3-body phase-space population have been identified in the kinematical distributions of $K^{*}$, as well as no convincing signal of a spin-alignment has been observed.

This measurement sets a baseline for future studies of $K^{*}$ production in proton-nucleus
and heavy-ion collisions. For instance, by comparing the present data with previously measured by HADES proton-niobium collisions at the same beam energy, cold nuclear matter effects affecting the production of the $K^{*}$'s can be extracted.

\begin{acknowledgments}
The HADES collaboration gratefully acknowledges the support by the grants LIP Coimbra, Coimbra (Portugal) PTDC/FIS/113339/2009; SIP JUC Cracow, Cracow (Poland): NCN Poland, 2013/10/M/ST2/00042; Helmholtz-Zentrum Dresden-Rossendorf (HZDR), Dresden (Germany) BMBF 05P12CRGHE; TU M\"unchen, Garching (Germany) MLL M\"unchen: DFG EClust 153, VH-NG-330 BMBF 06MT9156 TP5 GSI TMKrue 1012; NPI AS CR, Rez, Rez (Czech Republic) M100481202 and GACR 13-06759S; USC - S. de Compostela, Santiago de Compostela (Spain) CPAN:CSD2007-00042, Goethe University, Frankfurt (Germany): HA216/EMMI HIC for FAIR (LOEWE) BMBF:06FY9100I GSI F\&E EU Contract No. HP3-283286.
\end{acknowledgments}

\bibliography{Kstar_pp_final}

\end{document}